# Self-Propagating High-Temperature Synthesis of Boron Subphosphide, $B_{12}P_2$

V. A. Mukhanov, P. S. Sokolov, O. Brinza, D. Vrel, and V. L. Solozhenko[*]

*LSPM–CNRS, Université Paris Nord, 93430 Villetaneuse, France*

**Abstract** – Two new methods to produce nanopowders of $B_{12}P_2$ boron subphosphide by self-propagating high-temperature synthesis have been proposed. Bulk polycrystalline $B_{12}P_2$ with microhardness of $H_V$ = 35(3) GPa and stability in air up to 1300 K has been prepared by sintering these powders at 5.2 GPa and 2500 K.

Icosahedral boron subphosphide $B_{12}P_2$ is a promising superhard material with theoretical hardness of $H_V$ = 37 GPa [1] and high thermal and chemical stability [2]. It may be produced by three methods: (i) thermal decomposition of BP boron phosphide at temperatures above 1500 K in a reducing atmosphere [3], (ii) direct interaction between the elements at $T$ > 1600 K under argon pressure of 50 bar [4], and (iii) reaction between boron and phosphorus halogenides [5]. $B_{12}P_2$ single crystals can be grown both by crystallization from flux solutions [5] and gas-transport reactions [3, 5]. However, technologies based on the above methods are complex and labor intensive, which prevents practical applications of boron subphosphide.

Earlier we have developed a method to produce BP boron phosphide free of $B_{12}P_2$ impurity by interaction of boron phosphate and magnesium according to the

$$BPO_4 + 4Mg = BP + 4MgO \qquad (1)$$

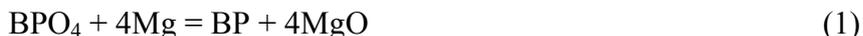

reaction in the mode of self-propagating high-temperature synthesis (SHS) [6]. As a result of side reactions and partial oxidation by air oxygen, the yield of the desired product (BP) is relatively low (~35% of the theoretical one), which, however, is compensated by the method simplicity and availability of the reagents used.

The aim of the present work was to develop efficient SHS methods of producing single-phase boron subphosphide by elaborating experimental approaches proposed in [6]. As a result of our studies two schemes of $B_{12}P_2$ synthesis were suggested.

**Scheme 1**. The synthesis was performed by changing the precursor stoichiometry, i.e. instead of $BPO_4$ we used a boron phosphate glass of $B_{12}P_2O_{23}$ composition produced by evaporation of a mixture of boric (Alfa Aesar, 99.8 %) and orthophosphoric (Alfa Aesar, 85% aq. sol.) acids in distilled water (3.2 : 1 : 2.1 weight ratio) at 520 K with subsequent annealing of the product in a muffle furnace at 770 K. The mixture of as-prepared $B_{12}P_2O_{23}$ glass (-200 μm) and metallic magnesium (Alfa Aesar, 99.8 %, 325 mesh) in the 1 : 1 weight ratio was pressed in a steel die at 10 t load into 2.2-g pellets 20 mm in diameter and 4 mm in height. The reaction was performed in a dynamic argon atmosphere (pressure of 1 bar, consumption of 50 cm$^3$/s) in the originally designed SHS apparatus described elsewhere [7]. The pellets placed on a substrate of pressed MgO were fired with a AC-heated graphite foil ribbon; the quantity of heat released on the

---

[*] e-mail: vladimir.solozhenko@univ-paris13.fr

ribbon was comparable with the thermal effect of the reaction (2). According to our estimations, the temperature in the course of the reaction was no less than 1300 K. After burning the weight of pellets decreased by ~6% (a part of the substance was taken away with the argon flow). The equation of the reaction may be written as

$$B_{12}P_2O_{23} + 23Mg = B_{12}P_2 + 23MgO \tag{2}$$

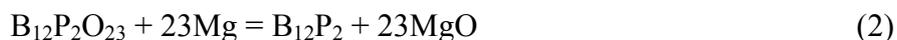

After boiling the resulting products in 20% hydrochloric acid for 30 min, there forms a brown residue, which is $B_{12}P_2$ containing no more than 3 vol % BP (Fig. 1a). The presence of a small BP impurity may be explained by the fact that the initial boron phosphate glass contained some amount of crystalline $BPO_4$. The yield of $B_{12}P_2$ made about 51% of the theoretical one according to reaction (2).

In the comparison experiments similar pellets were annealed at 870 K in a closed crucible in a muffle furnace for 20 min. After treating the reaction products with hydrochloric acid, the yield of $B_{12}P_2$ was 65% of theoretical by reaction (2). In this case, however, the BP impurity content increased up to ~10 vol%.

**Scheme 2**. The aim was achieved by the variation of the reducing agent stoichiometry during the reduction of boron phosphate by the reaction

$$2BPO_4 + 5MgB_2 + 3Mg = B_{12}P_2 + 8MgO \tag{3}$$

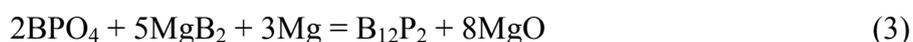

As a reducing agent the mixture of magnesium diboride (Alfa Aesar, 99 %, 100 mesh) and metallic magnesium (Alfa Aesar, 99.8 %; 325 mesh) in the weight ratio 3 : 1 was used. The SHS reaction was carried out according to the procedure similar to that used in scheme 1. A decrease in the pellets weights was ~3%. After boiling the products in 20% hydrochloric acid for 30 min, the light-grey residue was single-phase $B_{12}P_2$ (see Fig. 1b). The yield of boron subphosphide was 76% of theoretical by reaction (3). Similar pellets annealed in a muffle furnace in a closed crucible at 870 K for 20 min after washing in 20% hydrochloride acid allow us to produce $B_{12}P_2$ with a 71% yield, and in this case the BP content increases up to 16 vol%.

Relatively low (50–75%) yields of the desired product by reactions (2) and (3) are due to a number of side reactions (the formations of magnesium phosphide, diboride and borate; boron and phosphorus oxides, etc.), which, however, as in the case of BP [6], is compensated by the simplicity of the method and availability of the reagents used.

The X-ray diffraction analysis of the products was performed on an Equinox 1000 Inel diffractometer (CuKα$_1$ radiation, λ = 1.540598 Å). The lattice parameters of boron subphosphide samples prepared according to schemes 1 and 2 are $a$ = 5.985(3) Å, $c$ = 11.842(9) Å and $a$ = 5.988(3) Å, $c$ = 11.836(7) Å, respectively, which agrees well with the values reported for $B_{12}P_2$ with stoichiometry close to the ideal [4]. The sizes of coherent scattering areas (CSA) estimated from the diffraction lines broadening using Williamson–Hall method [8] were 20–30 nm regardless of the production scheme.

The morphology of $B_{12}P_2$ powders was studied on a Supra 40VP Carl Zeiss high-resolution scanning electron microscope. According to the data obtained, the $B_{12}P_2$ powders consist of isotropic grains with size of 50–90 nm irrespective of the production scheme (Fig. 2). A comparison of these values with the CSA sizes allows a conclusion that $B_{12}P_2$ grains observed in the electron microscope (see Fig. 2) are aggregates of several crystallites. The local element analysis of washed boron subphosphide powders was made on a Leica S440 electron microscope with an EDS Princeton Gamma-Tech energy-dispersive spectrometer. According to the data obtained, for all samples the B : P ratio is 6 : 1, and total impurities concentration (Mg, O, Cl, Al, and Si) does not exceed 0.7 at%.

The phase purity of the boron subphosphide samples was also verified by Raman scattering. Raman spectra were excited by a He-Ne laser ($\lambda$ = 632.8 nm, beam size = 10 μm) and recorded on a Horiba Jobin Yvon HR800 micro-Raman spectrometer. The Raman spectra of washed reaction products are shown in Fig. 3. All observed lines correspond to $B_{12}P_2$ [9] and only some spectra of samples prepared by scheme 1 exhibit a weak band at ~810 cm$^{-1}$ that is characteristic of BP [6] (see Fig. 3a).

Studies of chemical properties of the synthesized $B_{12}P_2$ and BP powders showed that at boiling they are stable to the action of both 30% hydrochloric and 30% nitric acids, and slowly dissolve in a mixture (3 : 1 volume ratio) of 37% HCl and 68% $HNO_3$, while boron and phosphorus (both red and black) dissolve in 20% nitric acid even at room temperature. At boiling of the mixture of $B_{12}P_2$ and BP powders (9 : 1 weight ratio) in 96% sulphuric acid for 30 min, approximately 17% by weight of the sample dissolves. In this case BP dissolves completely, and this allows us to use this procedure to remove small BP impurities from $B_{12}P_2$. In concentrated (> 80%) solutions of alkalis (NaOH, particularly) $B_{12}P_2$ and BP slowly dissolve at temperatures above 520 K.

The washed $B_{12}P_2$ powders were sintered at 5.2 GPa and 2500 K for 3 min in a high-temperature cell of a toroid-type high-pressure apparatus. The experimental details were described earlier [10]. The recovered samples were single-phase non-porous polycrystalline bulks of boron subphosphide with lattice parameters of $a$ = 5.992(9) Å, $c$ = 11.859(3) Å and grain sizes of 1–2 μm. The Vickers hardness of these bulks measured using a Duramin-20 (Struers) microhardness tester at loads up to 20 N and indentation time of 10 s was 35(3) GPa, which practically coincides with the theoretical value of $H_V$ = 37 GPa [1] calculated for $B_{12}P_2$ in the framework of the thermodynamic model of hardness [11].

The thermal stability of boron subphosphide in the 300–1500 K temperature range was studied by thermogravimetry (TG) and differential scanning calorimetry (DSC) on a Netzsch STA 409 PC thermal analyzer at continuous heating rate of 10 K/min in an air flow (30 cm$^3$/min) using flat-plate sample holders of aluminum oxide. The results of thermoanalytical studies of $B_{12}P_2$ nanopowder and bulk polycrystalline material produced by its sintering at high pressures and high temperatures are shown in Fig. 4. The nanodispersed boron subphosphide is stable in air up to 900 K, and at higher temperatures it oxidizes to $B_2O_3$ and $BPO_4$ that is accompanied by evaporation of a part of the forming boron (III) oxide. Under similar conditions $B_{12}P_2$ bulk material starts to oxidize only at temperatures above 1300 K.

CONCLUSIONS

Single-phase $B_{12}P_2$ boron subphosphide nanopowders have been produced by self-propagating high-temperature synthesis as a result of reduction of (i) boron phosphate by a mixture of $MgB_2$ and metallic magnesium, and (ii) boron phosphate glass of $B_{12}P_2O_{23}$ composition by metallic magnesium; with a subsequent chemical purification of the reactions products. The suggested methods are characterized by the simplicity of implementation, high efficiency, low cost of the product, and good perspectives for large-scale production. The synthesized $B_{12}P_2$ powders have been studied by powder X-ray diffraction, scanning electron microscopy, Raman spectroscopy, and thermal analysis. High-dense polycrystalline bulks with Vickers microhardness of 35(3) GPa and high (up to 1300 K) thermal stability in air have been produced by sintering $B_{12}P_2$ powders at 5.2 GPa and 2500 K.

ACKNOWLEDGEMENTS

The authors thank Drs. T. Chauveau, A. Tallaire, and T.B. Shatalova for their assistance in X-ray diffraction analysis, Raman spectroscopy studies, and thermal analysis. This work was financially supported by the Agence Nationale de la Recherche (grant ANR-2011-BS08-018) and DARPA (grant W31P4Q1210008).

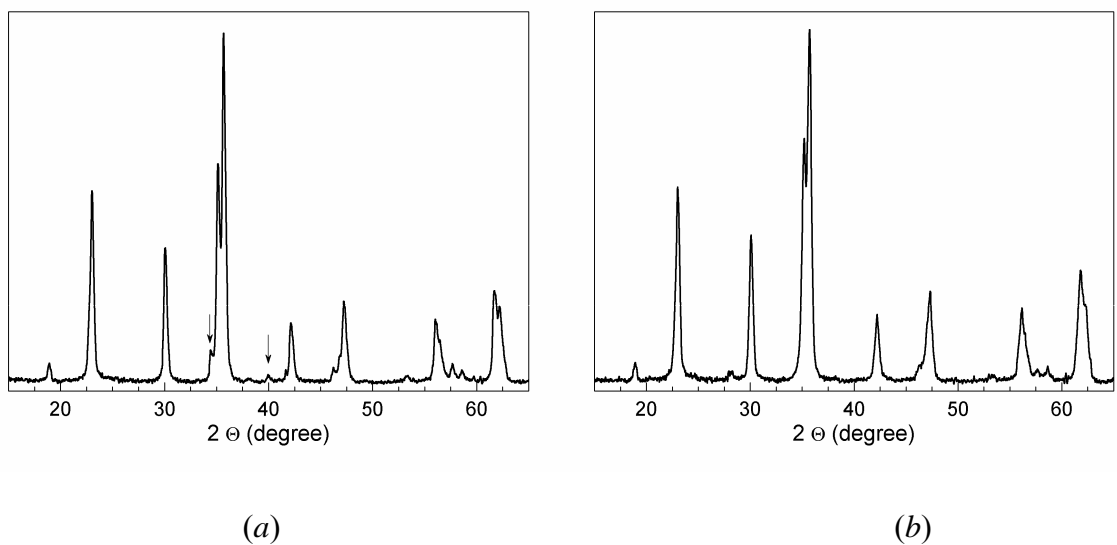

**Fig. 1.** Diffraction patterns of washed $B_{12}P_2$ samples produced by SHS according to schemes 1 (a) and 2 (b). Arrows show locations of 111 and 200 diffraction lines of BP.

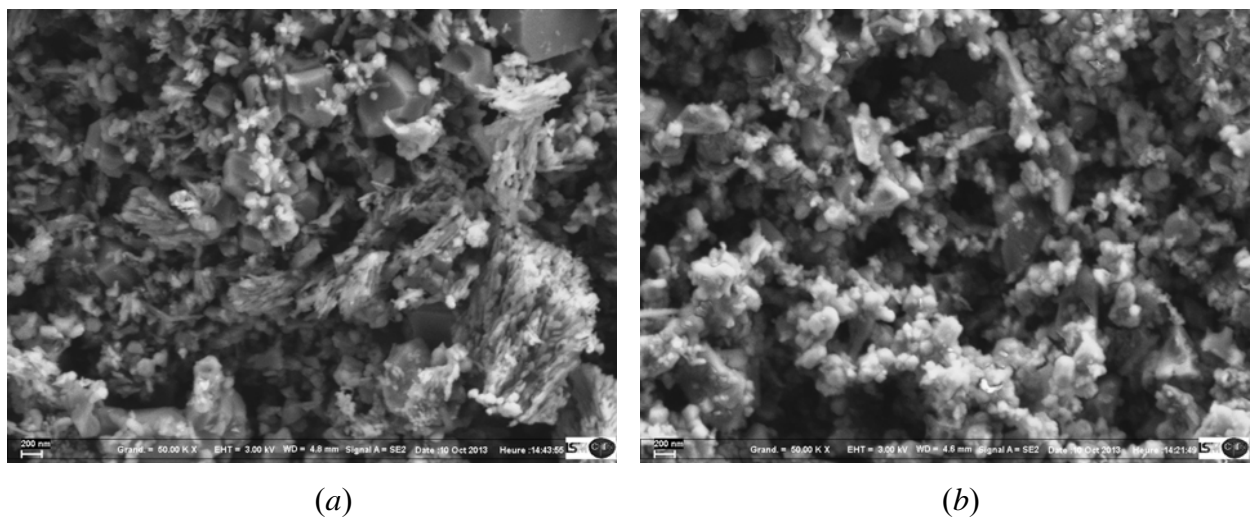

**Fig. 2.** SEM micrographs (50000×) of washed $B_{12}P_2$ samples produced by SHS according to schemes 1 (a) and 2 (b).

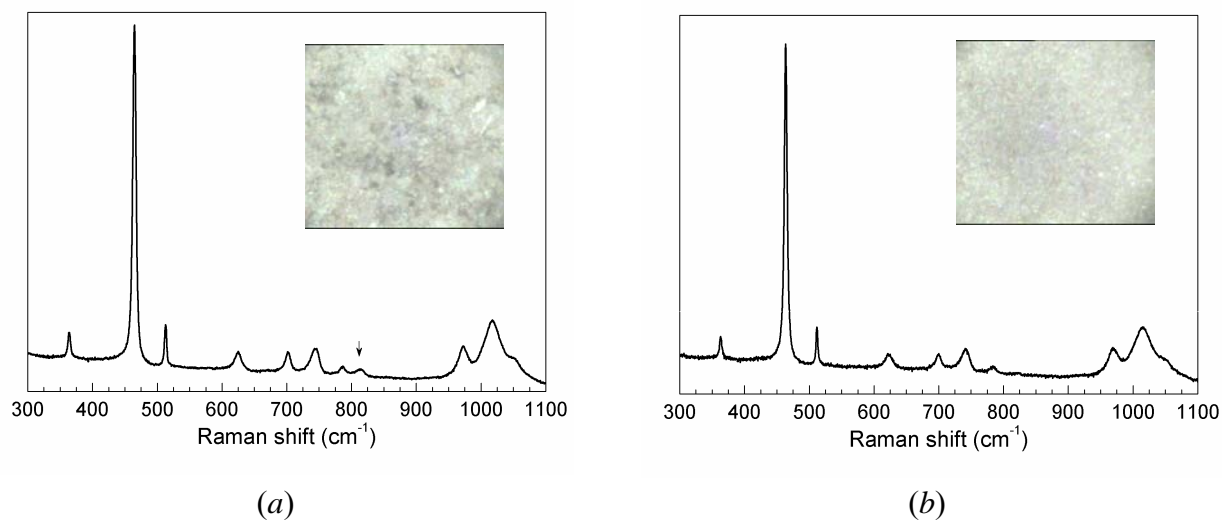

**Fig. 3.** Raman spectra of washed $B_{12}P_2$ samples produced by SHS according to schemes 1 (a) and 2 (b). The arrow points the location of the most intensive BP line (810 cm$^{-1}$); the inserts show optical images (100×) of the powders surfaces.

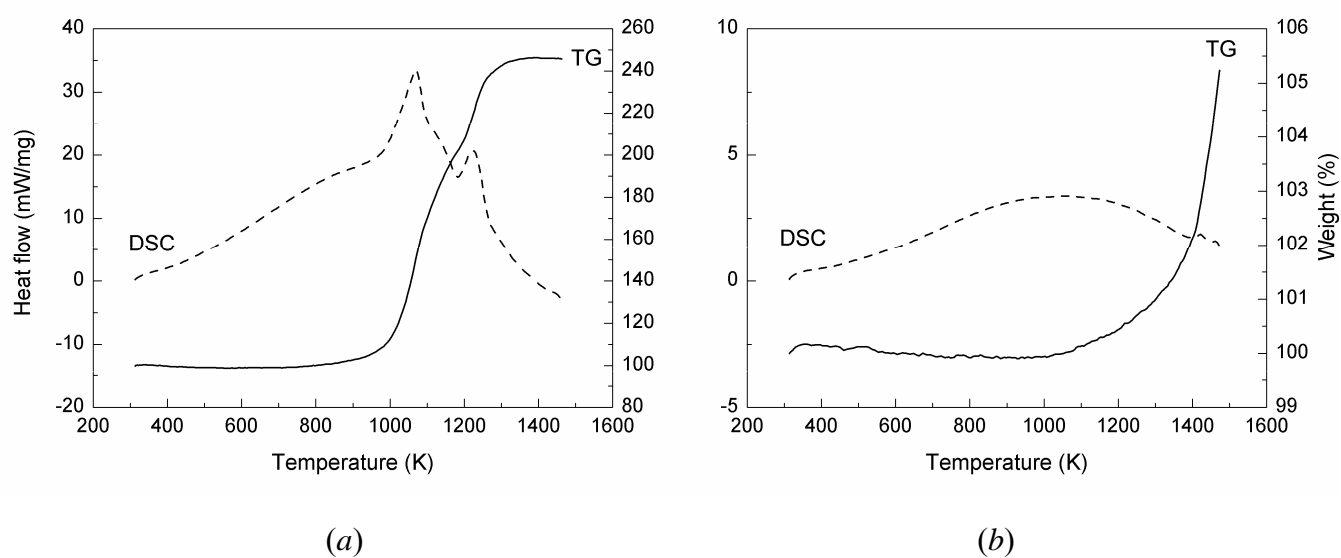

**Fig. 4.** TG and DSC curves of the $B_{12}P_2$ nanopowder synthesized by SHS using scheme 2 (a) and polycrystalline bulk $B_{12}P_2$ produced by sintering at 5.2 GPa and 2500 K (b).